\documentclass[twocolumn,aps,prc,showpacs,preprintnumbers,amsmath,amssymb,floatfix]{revtex4}
\usepackage{graphicx}
\usepackage{bm}
\usepackage{color}
\usepackage{hyperref}
\usepackage{caption}
\begin{document}
\title{Selection of special orientations in relativistic collisions of deformed heavy nuclei}
\author{C. Nepali}
\author{G. Fai}
\author{D. Keane}
\affiliation{Center for Nuclear Research, Department of Physics \\
	     Kent State University, Kent, OH 44242, USA}
\date{\today}
\begin{abstract}
We have studied U+U collisions at $\sqrt{s_{NN}}$ = 200 GeV using Monte Carlo Glauber, UrQMD and AMPT models. 
We find that it is possible to separate central tip-tip events as well as central body-body events 
on the basis of cuts on multiplicity and magnitude of the reduced flow vector.
\end{abstract}
\maketitle
%
%
\section{Introduction}
One of the recent surprises about the properties of the QCD matter formed in 
Au+Au collisions at RHIC is that it appears to flow with near-zero viscosity: an almost perfect
liquid~\cite{Kolb:2003dz,Heinz:2004qz,Kolb:2001qz,Adams:2005dq,Hirano:2005wx,Tannenbaum:2006ch}.
This is in contrast with prior expectations of non-interacting gas-like behavior, and is based in part
on approximate agreement between results from non-viscous hydrodynamic calculations and experimental data  
on elliptic flow. Azimuthal anisotropy in the transverse plane, or elliptic flow ($v_{2}$), is sensitive to the 
early pressure developed in the system and its equation of state. There is a close connection 
between elliptic flow and the initial spatial eccentricity ($\epsilon$) of the transverse overlap 
region of the two colliding nuclei. Non-viscous hydrodynamic calculations predict constant $v_2 / 
\epsilon$ over a wide range of impact parameters~\cite{Kolb:2003dz}. While the data remain well below 
the non-viscous hydrodynamic values at lower beam energies, central collisions of Au+Au at $\sqrt{s_{NN}}$ = 
200 GeV just reach the prediction~\cite{Adams:2005dq,Voloshin:2007af}. 

In relativistic uranium-uranium (U+U) collisions, there is the potential to produce more 
extreme conditions of excited matter (higher density and/or a greater volume of highly excited 
matter) than is possible using spherical nuclei like Au+Au or Pb+Pb at the same incident 
energy~\cite{Shuryak:1999by,Li:1999be,Heinz:2004ir,Kuhlman:2005ts,Nepali:2006ep}. However, 
this potential can be partly lost if we have limited capability to distinguish experimentally 
between different collision orientations when the ions interact near zero impact parameter.  
The prior studies cited above have devoted attention to this problem, and while there is 
agreement that U+U collisions offer worthwhile advantages over Au+Au, quantitative particulars 
have yet to be worked out in detail~\cite{Heinz:2004ir,Kuhlman:2005ts,Nepali:2006ep}.  
Furthermore, U+U collisions offer the opportunity to explore a different and wider range 
of initial eccentricities than is possible in the case of Au+Au.  
Among the possible collision orientations, of particular interest are the ``tip-tip" orientation, 
in which the long axes of both deformed nuclei are aligned with the beam axis, and the ``body-body" 
orientation, in which the long axes are both perpendicular to the beam axis and parallel to each 
other. In this paper, we report a study of the selection of these orientations based on a 
Monte-Carlo Glauber model~\cite{Glauber,Adams:2003yh}, Ultra-relativistic Quantum Molecular 
Dynamics (UrQMD 1.3)~\cite{Bass:1998ca}, and A Multi Phase Transport Model (AMPT 
v1.11-v2.11)~\cite{Lin:2004en}.

The Glauber model is solely based on collision geometry~\cite{Glauber}. 
The UrQMD model is a microscopic transport theory based on the covariant propagation of all hadrons 
on classical trajectories in combination with stochastic binary scatterings, color string formation 
and resonance decay~\cite{Bass:1998ca}. We use this model with default settings.
AMPT is a hybrid model. It uses the heavy-ion jet interaction generator (HIJING) to generate initial 
conditions, Zhang's parton cascade (ZPC) for the partonic scattering and hadronization, and a 
relativistic transport (ART) model for hadronic interactions and freeze-out~\cite{Lin:2004en}. 
We use this model with string melting, where all excited strings that are not projectile and target 
nucleons without any interactions are converted to partons according to the flavor and spin structures 
of their valence quarks. This option with appropriate partonic cross-section gives higher value of elliptic
flow, close to the experimental value, than without string melting~\cite{Lin:2004en}. We keep all other 
options at their default settings.

The article is organized as follows: in Sec.~\ref{sec:calculation}, we briefly review the relevant quantities 
and outline our calculation, and in Sec.~\ref{sec:selection}, we present the selection procedure for 
tip-tip and body-body events and discuss the results. In Sec.~\ref{sec:conclusion}, we summarize our findings.
\section{Calculational framework}
\label{sec:calculation}
We represent the quadrupole deformation of the ground-state uranium nucleus in the standard~\cite{Bohr:1969} 
way: we take a Saxon-Woods density distribution with surface thickness $a = 0.535$ fm and with 
$R = R_{\text {sp}}(0.91 + 0.27 \cos^2\theta)$, where $\theta$ is the polar angle relative to the long axis 
of the nucleus and $R_{\text sp} = 1.12 \; A^{1/3} - 0.86 \; A^{-1/3}$ fm~\cite{Bohr:1969}. The small 
hexadecapole moment of the uranium nucleus is neglected. This yields $R_{\text{long}}/R_{\text{perp}} \approx$ 
1.29~\cite{Nepali:2006ep}. The orientation of the first and second nucleus in the colliding pair is fixed by 
the two angles ($\theta_{p} , \phi_{p}$) and ($\theta_{t} , \phi_{t}$), respectively. 
The angles $\theta_{p}$ and $\theta_{t}$ describe the orientation of the long axis relative to the beam direction, 
and they are uniformly distributed in [0, $\pi$/2].  The azimuthal angles $\phi_{p}$ and $\phi_{t}$ describe 
rotations about the beam direction, and they are uniformly distributed in [0, 2$\pi$]. 
The Monte-Carlo Glauber simulation is as described in Ref.~\cite{Nepali:2006ep}, and charged multiplicity 
in the central pseudo-rapidity region is parametrized using the approach of Ref.~\cite{Kharzeev:2000ph}:  
\begin{equation} \label{equ:para}
\frac{dN_{\text{ch}}}{d\eta}\Bigr\rvert_{\eta=0} = n_{\text{pp}} \; [x \: N_{\text{b}} + (1-x) \: 
\frac{N_{\text {w}}}{2}] \: ,
\end{equation}
where $n_{\text{pp}}$ = 2.19 and $x$ = 0.15 are obtained by fitting the PHOBOS data \cite{Back:2002uc} for Au+Au 
at 200 GeV~\cite{Nepali:2006ep}.  Here, $N_{\text{b}}$ and $N_{\text{w}}$ are the number of binary collisions, 
and number of participant (wounded) nucleons, respectively.
The multiplicity is then distributed according to a negative binomial~\cite{Ansorge:1988kn} in order to 
get a realistic distribution. To convert between track densities in rapidity $y$, and pseudo-rapidity 
$\eta$, we use the approximation $dN_{\text{ch}}/dy \approx 1.15~ dN_{\text{ch}}/d\eta$~\cite{Adler:2002pu} 
in the Monte-Carlo Glauber calculations. The UrQMD and AMPT codes have been modified to simulate the deformed 
uranium nuclei. 

In this paper, ``ideal tip-tip'' and ``ideal body-body'' refer to the configurations in which the long axes
of both deformed nuclei are aligned with the beam axis at zero impact parameter 
($\theta_p = \theta_t = \phi_p = \phi_t = 0, \; b = 0$ fm) and in which the long axes are both perpendicular 
to the beam axis and parallel to each other at zero impact parameter 
($\theta_p = \theta_t = \pi/2$, $\phi_p = \phi_t = 0, \; b= 0$ fm), 
respectively~\cite{Li:1999be,Nepali:2006ep}. As the statistical weight of the above mentioned ideal 
configurations is negligible, ``tip-tip'' and ``body-body'' are used to refer to the configurations 
close to ``ideal tip-tip'' and ``ideal body-body'', respectively, for selection purposes.

In the standard way of calculating the spatial eccentricity of the transverse overlap region,
$\epsilon_{\rm std}$, the minor axis of the ellipse representing the transverse overlap region is
taken along the direction of the impact parameter.
However, the minor axis may not be along the direction of the impact parameter due to fluctuation 
in the participating nucleon positions~\cite{Manly:2005zy}. The spatial eccentricity calculated 
by taking into account the rotation of the minor axis is referred to as the participant eccentricity, 
$\epsilon_{\text{part}}$, and given by Ref.~\cite{Manly:2005zy}:
\begin{equation}\label{equ:epsilon}
	\epsilon_{\text{part}} = \frac{\sqrt{(\sigma^2_y - \sigma^2_x)^2 + 4\sigma^2_{xy}}}{\sigma^2_y + \sigma^2_x} \: ,
\end{equation}	
with
\begin{align}
	\sigma^2_x& = \langle x^2 \rangle - \langle x \rangle^2 \: , 
&	\sigma^2_y& = \langle y^2 \rangle - \langle y \rangle^2 \: ,\\ 
	\sigma_{xy}& = \langle xy \rangle - \langle x \rangle \langle y \rangle \: ,
\end{align}	
where $x$ and $y$ are the transverse coordinates of the participant nucleons and 
angle brackets denote averaging over participant nucleons in a single event.  
For spherical nuclei, the difference between $\epsilon_{\rm std}$ and $\epsilon_{\rm part}$ is significant 
only in smaller systems or smaller overlap regions~\cite{Manly:2005zy}. However, in case of collisions of 
deformed nuclei, the transverse overlap region may not be an ellipse depending on the collision configurations. 
Because of this, the difference between $\epsilon_{\rm std}$ and $\epsilon_{\rm part}$ is significant 
even in central collisions. We use $\epsilon_{\rm part}$ in our calculations.

\noindent The area of the transverse overlap region is given by
\begin{equation}\label{equ:area}
	S = \pi \sqrt{\sigma^2_x \: \sigma^2_y} \: .
\end{equation}	
The event flow vector for the second harmonic, $\vec{Q}_2$, is defined as~\cite{Poskanzer:1998yz}
\begin{equation}
\vec{Q}_2 = \sum_{\substack{k}}^{\substack{n_{\text ch}}} \; \hat{i} \, w_k \, \cos 2 \phi_k \: \: 
       + \: \sum_{\substack{k}}^{\substack{n_{\text ch}}} \; \hat{j} \, w_k \, \sin 2 \phi_k \: \: .
\end{equation}
We also use the corresponding reduced flow vector, whose magnitude is 
\begin{equation}\label{equ:q2}
	q_{2} = |\vec{Q}_{2}|/\sqrt{n_{\text ch} \langle w_k^2 \rangle} 
	      = |\vec{Q}_{2}|/\sqrt{\sum_{\substack{k}}^{\substack{n_{\text ch}}} \: w_k^2} \:\,\, ,	
\end{equation}
where $\phi_k$ is the azimuthal angle of a particle, $w_k$ is a weighting factor for that particle, and the 
summation is over all $n_{\text{ch}}$ charged particles in the event.  The magnitude of the reduced flow vector, 
$q_{2}$ as defined above, is used in this work to establish a measure of flow strength where the trivial 
dependence on $n_{\text{ch}}$ is removed~\cite{Adler:2002pu}. 
The transverse-momentum-integrated elliptic flow, $v_2$, is parameterized as 
$v_2 = 0.034 \: \epsilon_{\text{part}} \: (dN_{\text {ch}}/dy)^{1/3}$~\cite{Sorensen:2006nw}, 
where $dN_{\text{ch}}/dy$ is the central rapidity density of the
charged particles. This expression is known to give a good description of Au+Au collisions at 
$\sqrt{s_{NN}}$ = 200 GeV and we assume here that it also holds for U+U collisions.  
 
To include $v_2$ in the Glauber model, azimuthal angles $\phi$ for 2$(dN_{\text{ch}}/d\eta)$ tracks are
generated according to the above parametrization, and we take $w_k = 1$.  In the 
UrQMD and AMPT models, $w_k = p^\perp_k$ for $p^\perp_k \leq 2.0$ GeV$/c$ and $w_k = 2.0$ GeV$/c$ 
otherwise, where $p^\perp_k$ is the transverse momentum of the $k^{\text{th}}$ particle in the event.

The quantities $S$ and $\epsilon_{\text{part}}$ for all the above-mentioned models were 
calculated on the basis of the Glauber model.  
The transverse particle densities, $(1/S)(dN_{\text{ch}}/dy)$~\cite{Kuhlman:2005ts,Nepali:2006ep} 
for ideal cases are:
\begin{equation}\label{equ:density}
\frac{1}{S}\frac{dN_{\text{ch}}}{dy} \Bigr\rvert_{y=0}	\approx	\begin{cases}	
									43.9 \: \text{fm}^{-2}&	\qquad \text{ideal tip-tip} \:, \\
									31.0  \: \text{fm}^{-2}&	\qquad \text{ideal body-body} \:.
								\end{cases}		
\end{equation}
%
%
\section{Selection Procedure}
\label{sec:selection}
The calculations based on the models show that the U+U central tip-tip configuration 
has the smallest overlap area, similar to central Au+Au, and results in the highest multiplicity.  
It is a good candidate to create the highest possible $(1/S)(dN_{\text{ch}}/dy)$ at any given beam energy. 
The higher the multiplicity, the better the applicability of hydrodynamics calculations~\cite{Kolb:2000sd}. 
The central body-body configuration also yields higher multiplicity than average, but smaller 
$(1/S)(dN_{\text{ch}}/dy)$, similar to central Au+Au, due to its large overlap area.  
However, it produces larger elliptic flow at the same $(1/S)(dN_{\text{ch}}/dy)$.
The tip-tip configuration yields higher $(1/S)(dN_{\text{ch}}/dy)$ at similar $\epsilon$, and the 
body-body configuration yields higher $\epsilon$ at similar $(1/S)(dN_{\text{ch}}/dy)$, compared 
to Au+Au at the same beam energy.
For comparison of the selected events with the ideal cases, we fitted the $S$ and $(1/S)(dN_{ch}/dy)$ 
distributions for ideal tip-tip and ideal body-body configurations with Gaussians. The quantity $q_2$ 
for ideal tip-tip is fitted according to~\cite{Adler:2002pu}: 
\begin{equation}
	\frac{dP}{q_2 \; dq_2} \propto \frac{1}{\sigma^2_2} \; \exp\left(-\frac{v^2_­2 M + q^2_2}{2\sigma^2_2}\right) 
	I_{0}\left(\frac{q_{2}v_{2}\sqrt{M}}{\sigma^{2}_{2}}\right) \: ,
\end{equation}	
with
\begin{equation}
	\sigma^{2}_{2} = 0.5 \; (1 + g_{2}) \: ,
\end{equation}
where $I_{0}$ is the modified Bessel function, $M = dN_{ch}/d\eta$, and $v_{2}$ is the elliptic flow.
The quantities $v_2$ and $g_{2}$ are taken as free parameters in fitting. The participant eccentricity, 
$\epsilon_{\rm part}$, is fitted similarly with different values of $g_2$.
For ideal body-body, a Gaussian fits well.

The following characteristics of the tip-tip and body-body configurations are used for selection:
\begin{equation}
\text{tip-tip}\Longrightarrow \begin{cases}	
					\text{largest} \; dN_{ch}/d\eta \; \text{(central collisions) \:,} \\
					\text{smallest} \; q_2 \, \text{, i.e., smallest}\; \epsilon_{\rm part}, \; \text{and}\\
					\text{smallest} \; S \: ;
	                        \end{cases}
\end{equation}
and 
\begin{equation}
\text{body-body}\Longrightarrow \begin{cases}
					  \text{largest} \; dN_{ch}/d\eta \; \text{(central collisions) \:,} \\
					  \text{largest} \; q_2 \, \text{, i.e., largest}\; \epsilon_{\rm part}, \; \text{and}\\
					  \text{largest} \; S \: .
					  \end{cases}
\end{equation}

The selected events are compared to the ideal cases. While $S$ and $\epsilon_{\rm part}$ cannot 
be directly measured, they are compared to the ideal cases for identification purposes.
\subsection{Tip-tip}
The quantity $N_{\text{w}}$ is approximately the same in tip-tip and body-body configurations because 
both have full overlap. However, $N_{\text{b}}$ is larger in tip-tip than body-body. Because of 
the contribution from $N_{\text{b}}$, the multiplicity is larger in tip-tip than body-body 
collisions. Thus, it is possible to select tip-tip on the basis of multiplicity only. However, 
for a variety of reason, this approach has relatively poor efficiency, and instead we have used 
both multiplicity and the reduced flow vector. 

\begin{figure}[!t]
		\caption*{Event distribution for tip-tip selection: Glauber}
		\resizebox{8.5cm}{!}{\includegraphics*{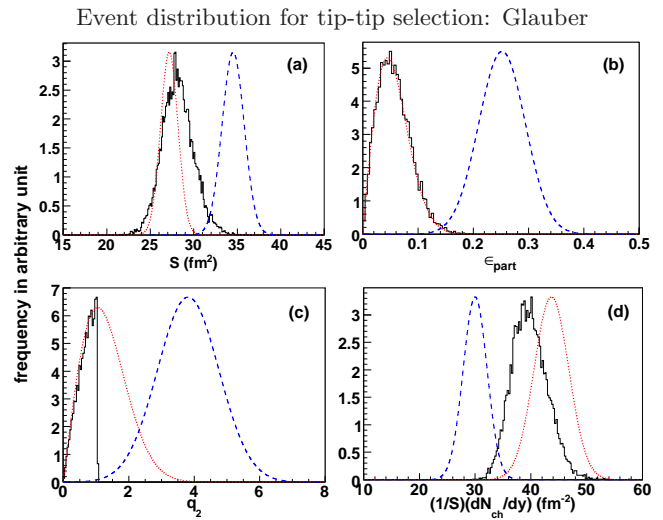}}
		\caption{
		(Color online) Distribution of (a) transverse overlap area, (b) eccentricity of the 
		transverse overlap region, (c) magnitude of the reduced flow vector, and (d) transverse 
		particle density from Monte-Carlo Glauber. The y-axes have arbitrary units. The dotted 
		and dashed curves are fitted for ideal tip-tip and ideal body-body configurations respectively, 
		compared here with the selected tip-tip events (solid lines).}
		\label{fig:tipSelectionGL}
\end{figure}

\begin{figure}[!t]
		\caption*{Event distribution for tip-tip selection: UrQMD}
		\resizebox{8.5cm}{!}{\includegraphics*{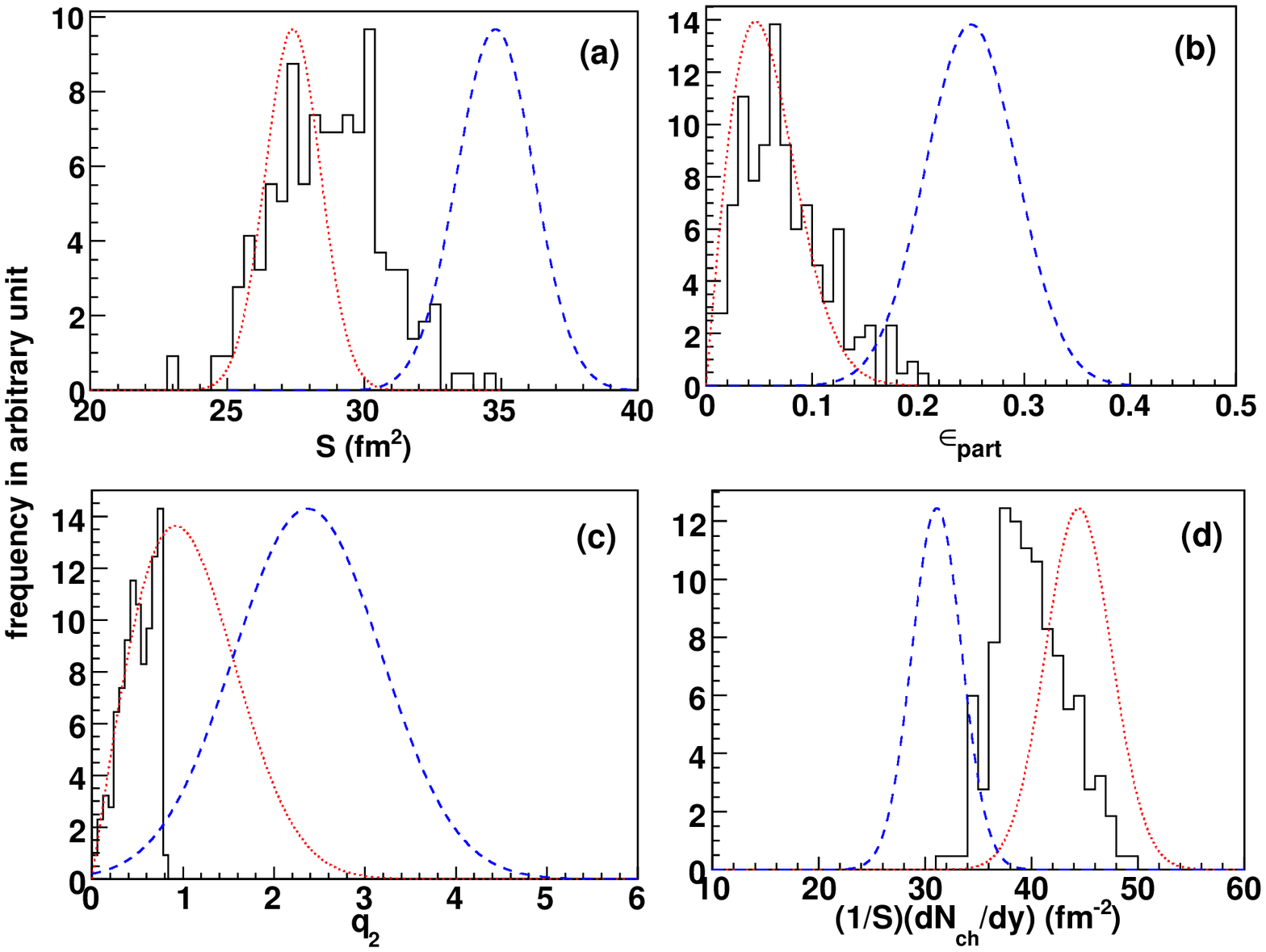}}
		\caption{
		(Color online) Same as Fig.~\ref{fig:tipSelectionGL} for UrQMD.}
		\label{fig:tipSelectionUR}
\end{figure}

\begin{figure}[!h]
		\caption*{Event distribution for tip-tip selection: AMPT}
		\resizebox{8.5cm}{!}{\includegraphics*{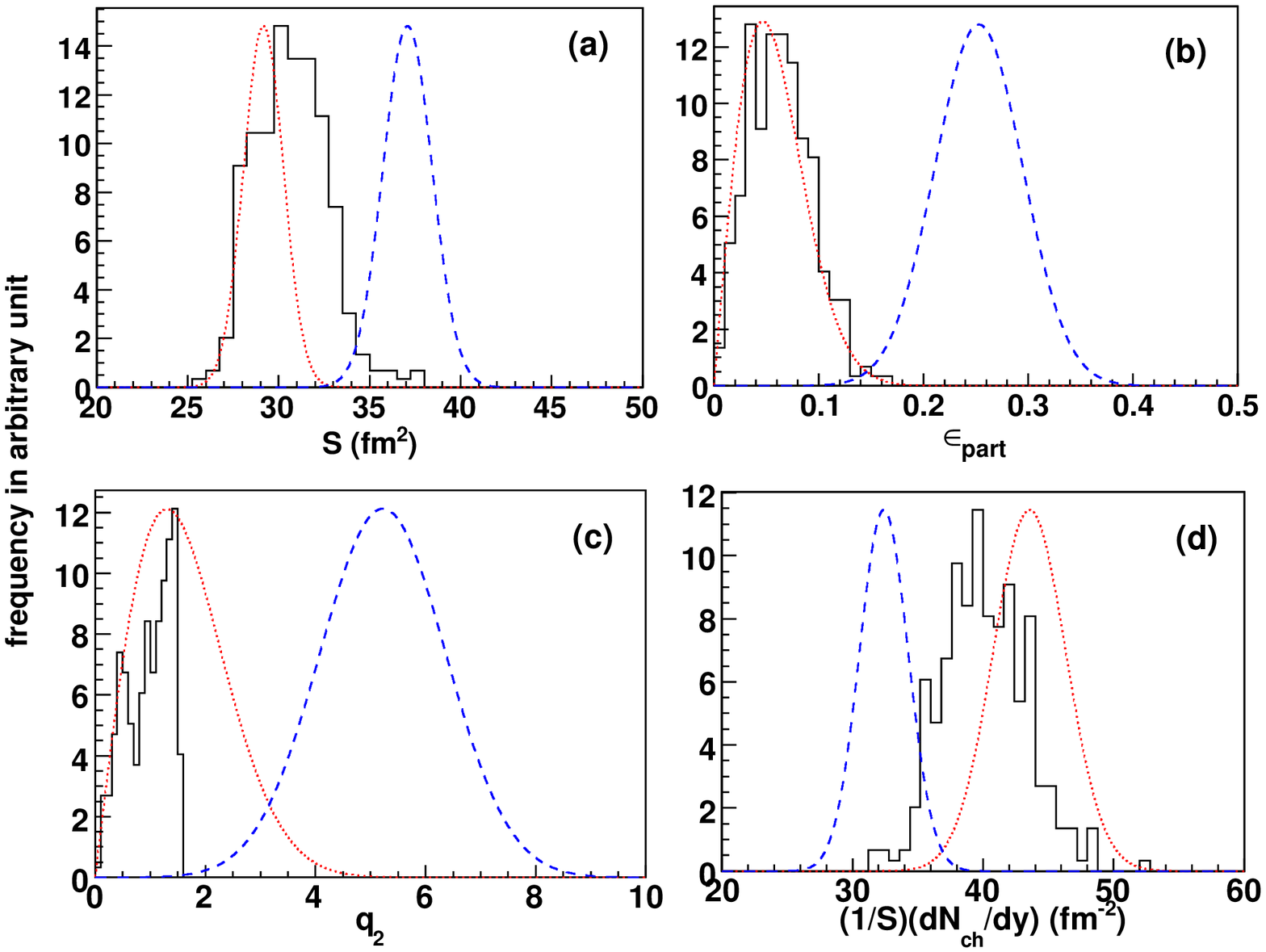}}
		\caption{
		(Color online) Same as Fig.~\ref{fig:tipSelectionGL} for AMPT.}
		\label{fig:tipSelectionAM}
\end{figure}

To select tip-tip events, first we take the top 1\% $dN_{\text{ch}}/d\eta$ to select
full overlap configurations. These are already rich in tip-tip, but a further selection of the bottom 
25\% $q_{2}$ is taken from the above subset to further enhance the tip-tip configuration. 
Figs.~\ref{fig:tipSelectionGL},~\ref{fig:tipSelectionUR}, and~\ref{fig:tipSelectionAM}~show the event 
distributions for the above three models, after cuts, compared with the ideal tip-tip case. 
The transverse particle density, $(1/S)(dN_{\text{ch}}/dy)$, obtained after these cuts is 
$\approx$ 40 $\text{fm}^{-2}$, which is significantly larger than Au+Au at the same beam energy.
\subsection{Body-body}
\begin{figure}[!h]
		\caption*{Event distribution for body-body selection: Glauber}
		\resizebox{8.5cm}{!}{\includegraphics*{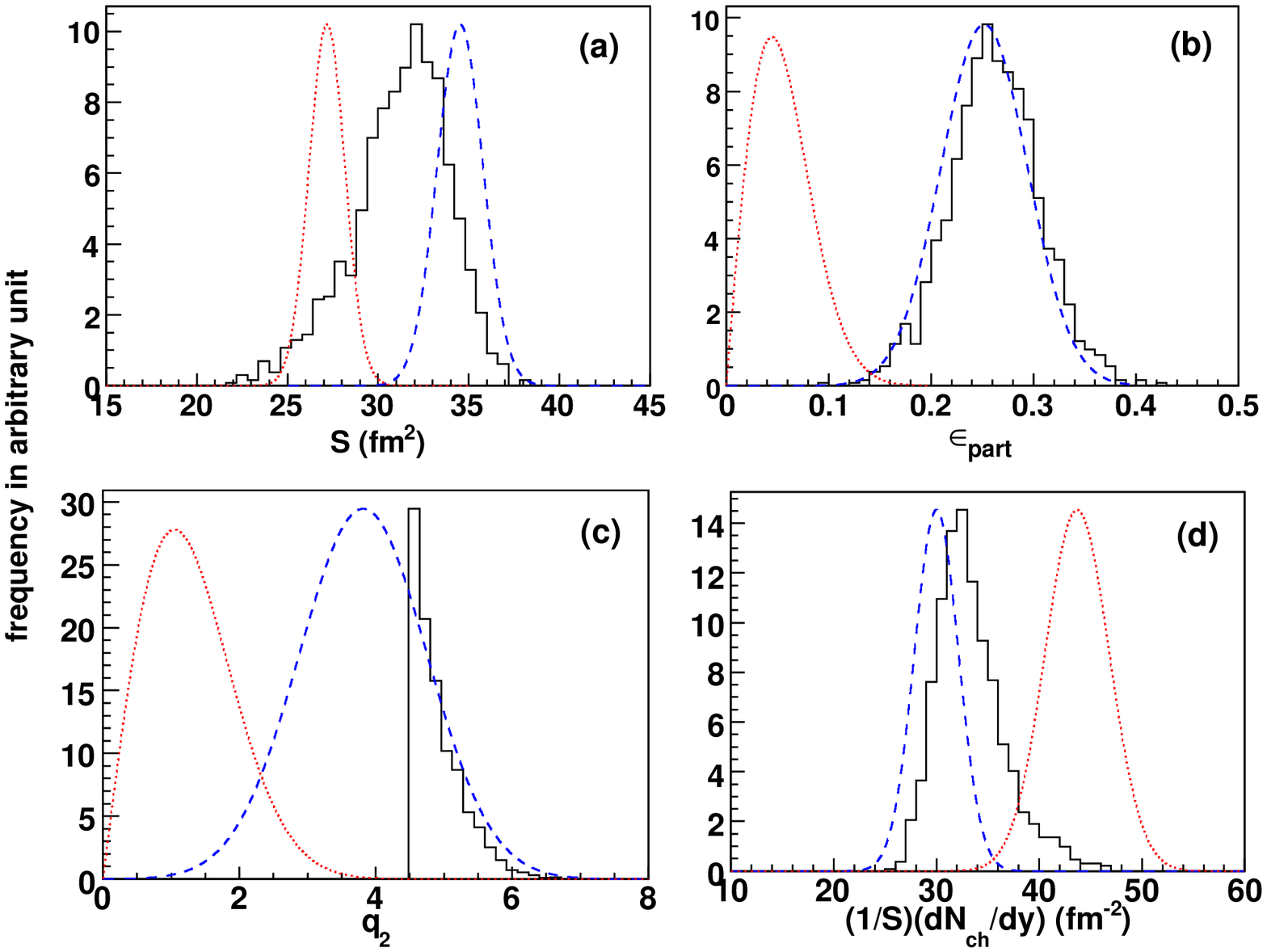}}
		\caption{
		(Color online) Same as Fig.~\ref{fig:tipSelectionGL} for selected body-body events in the Glauber model.}
		\label{fig:bodySelectionGL}
\end{figure}

\begin{figure}[!h]
		\caption*{Event distribution for body-body selection: UrQMD}
		\resizebox{8.5cm}{!}{\includegraphics*{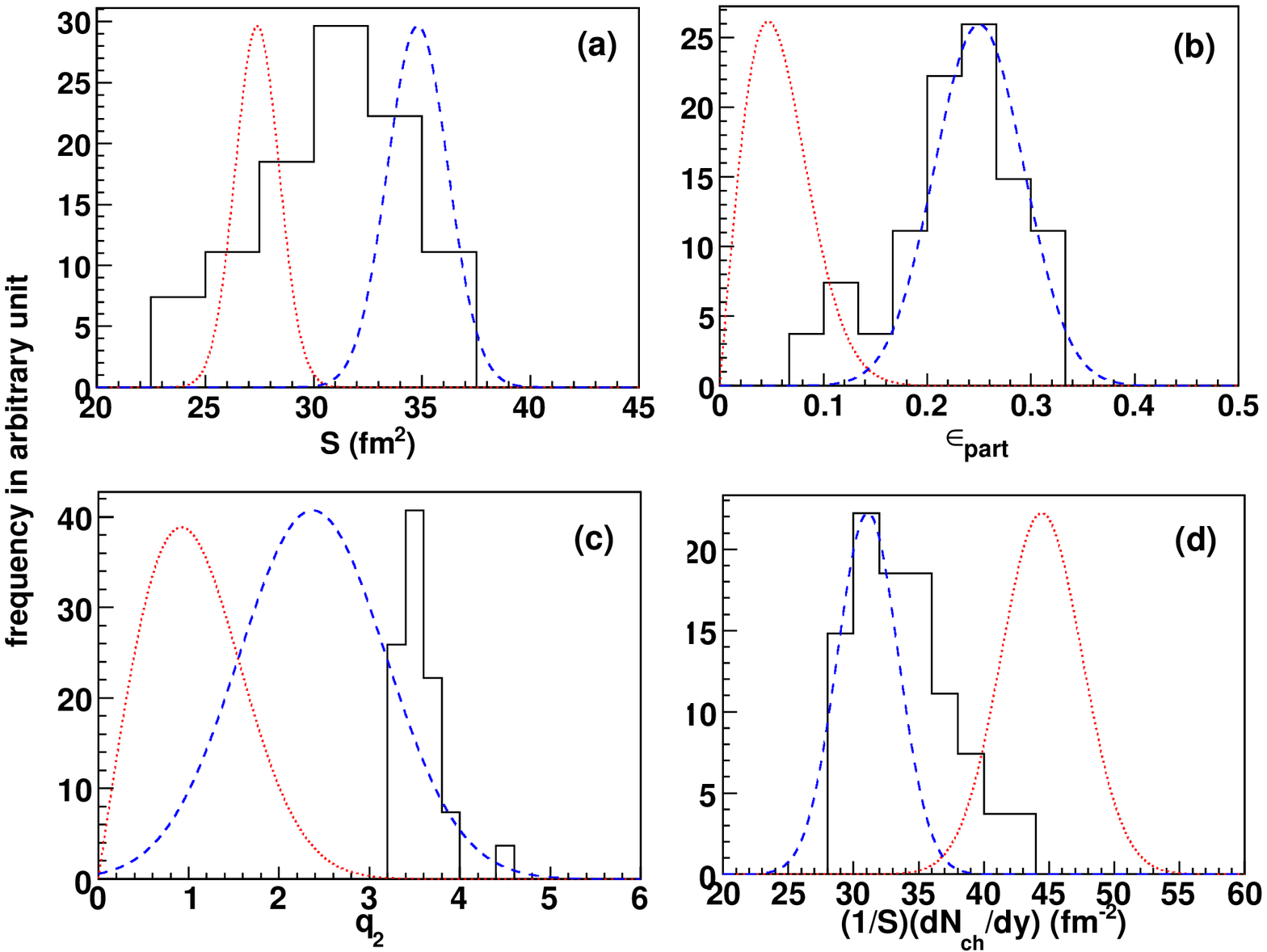}}
		\caption{
		(Color online) Same as Fig.~\ref{fig:tipSelectionGL} for selected body-body events in the UrQMD model.}
		\label{fig:bodySelectionUR}
\end{figure}

\begin{figure}[!h]
		\caption*{Event distribution for body-body selection: AMPT}
		\resizebox{8.5cm}{!}{\includegraphics*{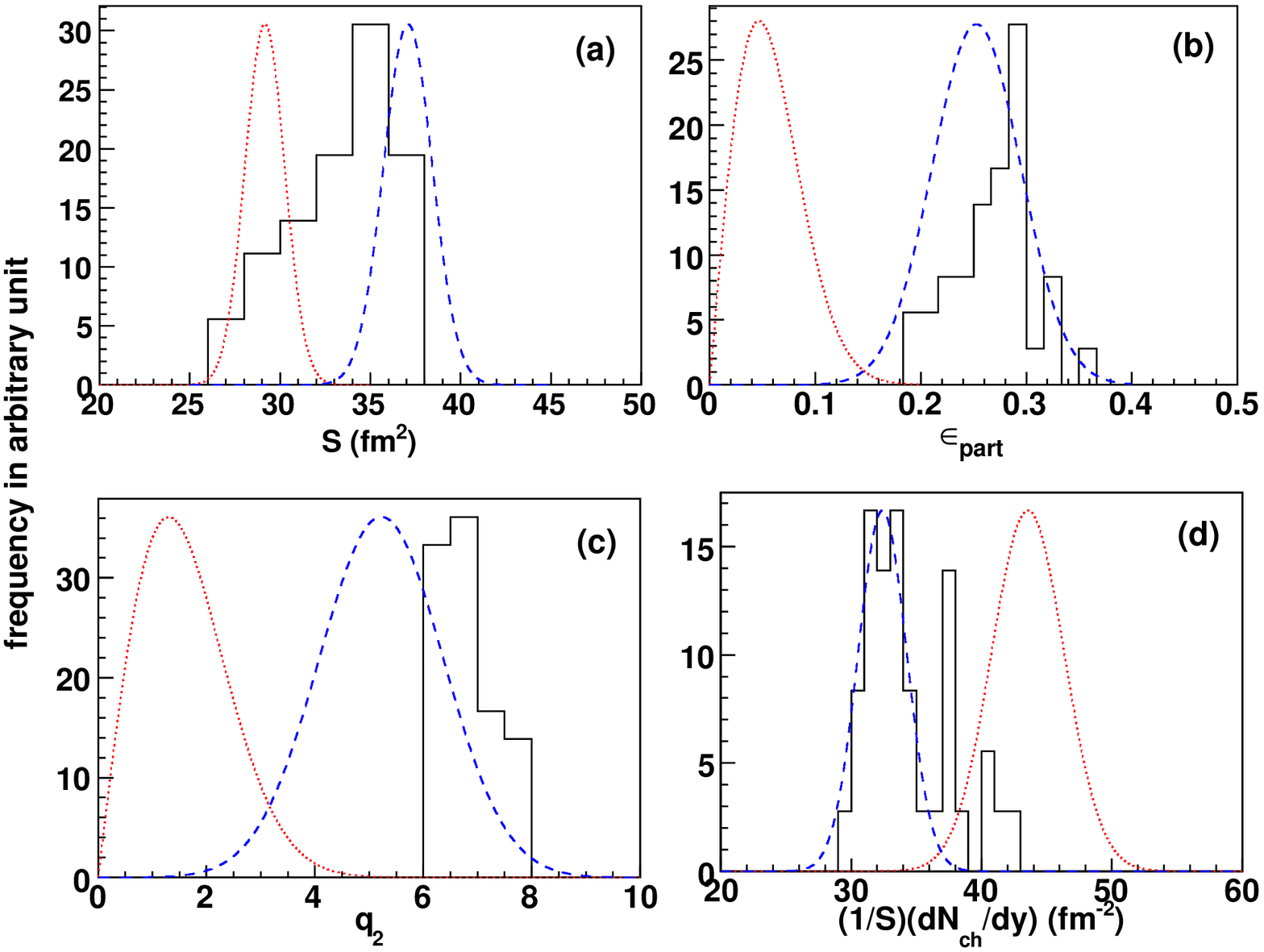}}
		\caption{
		(Color online) Same as Fig.~\ref{fig:tipSelectionGL} for selected body-body events in the AMPT model.}
		\label{fig:bodySelectionAM}
\end{figure}

Here, we first take the top 3\% $dN_{\text{ch}}/d\eta$ to select full overlap configurations. Then we
take the top 1\% $q_{2}$ to select body-body events. The $(1/S)(dN_{\text{ch}}/dy)$ obtained after all these 
cuts is $\approx$ 33 $\text{fm}^{-2}$. Figs.~\ref{fig:bodySelectionGL},~\ref{fig:bodySelectionUR}, 
and~\ref{fig:bodySelectionAM}~show the event distributions, after cuts, 
compared with the ideal body-body case, for the three models.

We have also tested an alternative procedure: take the top 3\% $dN_{ch}/d\eta$ to select full overlap
events. Then, take the lower 50\% $dN_{ch}/d\eta$ from the above subset to select mostly body-body events. 
This sample is further purified by taking the top 5\% $q_2$. However, 
this method is less efficient than the one discussed above.
\subsection{Discussion}
The center of the $S$ distribution in the ideal tip-tip and the ideal body-body from AMPT 
(Fig.~\ref{fig:tipSelectionAM}(a) and Fig.~\ref{fig:bodySelectionAM}(a)) differs from that for the Glauber model
(Fig.~\ref{fig:tipSelectionGL}(a) and Fig.~\ref{fig:bodySelectionGL}(a)) and for UrQMD 
(Fig.~\ref{fig:tipSelectionUR}(a) and Fig.~\ref{fig:bodySelectionUR}(a)) because of a different parametrization 
of the nuclear radius used in AMPT compared with the other two models. 
The $\epsilon_{\rm part}$ distributions for the selected events from all models are in good agreement with the 
ideal cases for both tip-tip (Fig.~\ref{fig:tipSelectionGL}(b), Fig.~\ref{fig:tipSelectionUR}(b), 
Fig.~\ref{fig:tipSelectionAM}(b)) and body-body (Fig.~\ref{fig:bodySelectionGL}(b), 
Fig.~\ref{fig:bodySelectionUR}(b), Fig.~\ref{fig:bodySelectionAM}(b)) selections. The quantity $q_2$ is  
sensitive to $\epsilon_{\rm part}$ and because of this, the hard cuts on $q_2$ bring the selected sample 
closer to the ideal values. 

The $dN_{ch}/d\eta$ cut selects mostly tip-tip events, with some body-body as well as other event categories 
having smaller $\epsilon_{\rm part}$ but larger $S$, depending on the orientation of the nuclei.  The 
additional cut in $q_2$ helps to remove most of the body-body events, but still leaves some events with smaller $\epsilon_{\rm part}$ but larger $S$. That is why the $S$ distribution for the selected events is a little 
different from the ideal cases.  The selection purity can be further improved by applying an even harder cut 
in $dN_{ch}/d\eta$.  A similar argument is also applicable to the body-body selection.
%
\section{Summary and Conclusion}
\label{sec:conclusion}
U+U collisions at maximum RHIC energy are predicted by three rather different
models to produce more extreme excitation of nuclear matter than has been achieved to date
using Au+Au.  The expected benefits of colliding U+U include a closer approach to the
conditions where hydrodynamics should be applicable, and the opportunity to explore a
different and wider range of initial eccentricities than is possible in the case of Au+Au. 
However, we can achieve the full potential benefits of U+U collisions only if we are able
to select favored collision orientations. 

The parameter space explored by colliding deformed nuclei is 
more varied than for spherical nuclei.
The tip-tip U+U configuration offers higher $(1/S)(dN_{\text{ch}}/dy)$, and lower
$v_2$ (similar to central Au+Au).  The body-body configuration gives higher $v_2$, and lower
$(1/S)(dN_{\text{ch}}/dy)$ (similar to central Au+Au).

We have presented a prescription for selection of tip-tip and
body-body configurations, based on the experimental observables $dN_{ch}/d\eta$ and $q_{2}$.
The $\epsilon_{\rm part}$ values for the selected events are in very good agreement with the ideal
cases in all models, although $S$ is somewhat different.
In our previous study~\cite{Nepali:2006ep}, we have estimated the benefits of U+U collisions
including the effect of detector resolution.
Here we have used a wider set of kinematic variables to demonstrate that the
desired configurations can be selected efficiently. With the development of the Electron
Beam Ion Source (EBIS)~\cite{Beebe:2000tz}, we expect U+U collisions at RHIC in the not-too-distant
future.

\section{Acknowledgment}
We thank Paul Sorensen for helpful ideas and discussions. We also thank Peter Jacobs for useful discussions.
This work was supported in part by U.S. DOE grants DE-FG02-86ER40251 and DE-FG02-89ER40531.
%
%

\end{document}